\documentclass{ifacconf}

\usepackage{amsmath}
\usepackage{amsfonts}
\usepackage{amssymb}
\usepackage{mathrsfs}
\usepackage{booktabs}
\usepackage{graphicx}
\usepackage{epstopdf}
\usepackage{epsfig}
\usepackage{color}
\usepackage{enumerate}
\usepackage{natbib}

\usepackage{tikz}
\usetikzlibrary{automata,arrows,positioning,calc}

\newtheorem{theorem}{Theorem}[section]
\newtheorem{lemma}[theorem]{Lemma}

\newtheorem{remark}[theorem]{Remark}
\newtheorem{problem}[theorem]{Problem}

\DeclareMathOperator{\Tr}{Tr}

\makeatletter

\newcommand{\Rmnum}[1]{\expandafter\@slowromancap\romannumeral #1@}
\makeatother

\begin{document}

\begin{frontmatter}

\title{\LARGE \bf
Optimal Scheduling of Multiple~Sensors with\\ Packet Length Constraint
}

\thanks[footnoteinfo]{The work by S. Wu and L. Shi is supported by an RGC General Research Fund 16210015}

\author[First]{Shuang Wu}
\author[First]{Xiaoqiang Ren}
\author[Third]{Subhrakanti Dey}
\author[First]{Ling Shi}

\address[First]{Department of Electronic and Computer Engineering, Hong Kong University of Science and Technology, Hong Kong, (e-mail: {swuak, xren, eesling@ust.hk}).}
\address[Third]{Department of Engineering Sciences, Uppsala University, Uppsala, Sweden, (e-mail: {Subhra.Dey@signal.uu.se})}

\begin{abstract}                % Abstract of not more than 250 words.
This paper considers the problem of sensory data scheduling of multiple processes. There are $n$ independent linear time-invariant processes and a remote estimator monitoring all the processes. Each process is measured by a sensor, which sends its local state estimate to the remote estimator. The sizes of the packets are different due to different dimensions of each process, and thus it may take different lengths of time steps for the sensors to send their data. Because of bandwidth limitation, only a portion of all the sensors are allowed to transmit in each time step. Our goal is to minimize the average of estimation error covariance of the whole system at the remote estimator. The problem is formulated as a Markov decision process (MDP) with average cost over an infinite time horizon. We prove the existence of a deterministic and stationary policy for the problem. We also find that the optimal policy has a consistent behavior and threshold type structure. A numerical example is provided to illustrate our main results.
\end{abstract}

\begin{keyword}
Sensor scheduling, Markov decision process, packet length, threshold policy.
\end{keyword}

\end{frontmatter}

%%%%%%%%%%%%%%%%%%%%%%%%%%%%%%%%%%%%%%%%%%%%%%%%%%%%%%%%%%%%%%%%%%%%%%%%%%%%%%%%%%%
\section{INTRODUCTION}

Wireless sensor networks are nowadays ubiquitous in various fields, e.g., industrial automation, habitat monitoring, smart grid and autonomous traffic management \citep{akyildiz2002wireless}. In typical settings, after taking local measurements, the sensors send their data to a remote estimator through a wireless channel according to a protocol. The wireless connection saves wiring and enables remote operation. The channel capacity of wireless communication, however, is often limited. What is more, the wireless sensor nodes are usually battery powered. In order to tradeoff the estimation quality and limited available resources, e.g., bandwidth and energy, it is critical to wisely allocate resources for the sensors.

To deal with the scheduling problem in the wireless sensor networks, researchers considered the constraints on the channel bandwidth, energy budget, etc. To maximize lifetime of the network, \citet{chamam2007optimal} studied a policy switching between active and inactive modes. Performance optimization under energy budget constraint has also been investigated. \citet{shi2011sensor} considered optimal sensor data scheduling problem for a single sensor under a limited energy budget. The authors proved the optimality of a periodic scheduling policy. In \cite{wu2013event}, an event-triggered transmission strategy was adopted to reduce the communication rate so as to save energy. In some applications, the sensors can interfere with each other, thus only a few sensors are allowed to be activated in each time slot. The sensor selection problem in this setting is considered, for example, in \cite{gupta2006stochastic}. Meanwhile, because of bandwidth limitation of communication channel, only a limited number of devices can transmit data at the same time. Optimal scheduling under channel bandwidth constraint was studied in \cite{zhao2014optimal}, \cite{mo2014infinite}, \cite{hovareshti2007sensor} and \cite{han2017optimal}. \cite{zhao2014optimal} and \cite{mo2014infinite} considered scheduling over an infinite horizon and proved that the optimal scheduling scheme can be approximated arbitrarily by a periodic schedule. \cite{hovareshti2007sensor} showed that the estimation quality can be improved for smart sensors equipped with some memory and processing capacities. \cite{han2017optimal} proved that the optimal scheduling policy over an infinite horizon can also be arbitrarily approximated by a periodic schedule for the smart sensor case.

The previous works assumed that it take the same time for transmission of packets in different processes. The transmission of one packet is assumed to be done in one time step. However, the scales of processes vary and the packets transmitted by different sensors may also be different. In typical communication protocols, e.g., Time Slotted Channel Hopping (TSCH, IEEE $802.15.4$e), the transmission duration of different packets can be different and the data transmission scheme can be designed by the engineers. In \cite{palattella2013optimal}, it was shown that TSCH protocol has superior energy efficiency compared with traditional IEEE802.15.4/ZigBee approaches. Standards of TSCH application can be found in \cite{watteyne2015using}.

Different from previous studies, our work focuses on the constraint of packet length. We study the sensor scheduling for remote state estimation of multiple LTI processes, each driven by white Gaussian noises, over an infinite horizon. Each process is measured by a sensor, which computes the local estimate of the process and sends the local estimate to a remote estimator. As the packet length of each process may be different, it costs different time duration for each sensor to complete one transmission of an estimate. Moreover, wireless sensor network has a bandwidth limitation: only a portion of all the sensor are allowed to transmit data to the remote estimator in one time step. We are interested in minimizing the average estimation error covariance of the remote estimator in this system. Our contributions are twofold.

1) We formulate the optimal scheduling of sensors under the packet length and bandwidth constraint as an average cost over an infinite time horizon Markov decision process. We prove that there exists a deterministic and stationary optimal policy by showing that the state space of Markov chain under optimal policy is finite.

2) By adopting dynamic programming, we prove that the optimal policy has a nice structure, i.e., consistency and threshold type structure. The consistency means that once a sensor is chosen to schedule, the transmission of the current estimate should not be interrupted by selecting another sensor. The threshold structure reduces searching space for off-line solution and facilitates online implementation.

The remainder of this paper is organized as follows. In Section~2, we develop the mathematical formulation of the problem of interest. The main results, which consist of the MDP formulation, the existence of a deterministic and stationary policy, and the consistency and the threshold structure of the policy, are given in Section~3. In Section~4, a numerical example is provided to illustrate the main results. We summarize the paper in Section~5.

\emph{Notation}: Denote $\mathbb{N}$ as the set of nonnegative integer numbers. For a matrix $X$, let $\Tr(X)$, $X^T$ and $\rho(X)$ represent the trace, the transpose and the spectral radius of $X$, respectively. The identity matrix is $I$, and its size is determined from the context. Let $P(\cdot)$ and $P(\cdot|\cdot)$ stand for the probability and conditional probability for certain events. Denote $\mathbb{E}[\cdot]$ as the expectation of a random variable.

%%%%%%%%%%%%%%%%%%%%%%%%%%%%%%%%%%%%%%%%%%%%%%%%%%%%%%%%%%%%%%%%%%%%%%%%%%%%%%%%%%%
\section{SYSTEM SETUP AND PROBLEM FORMULATION}
Consider the following $n$ independent discrete-time linear dynamic processes measured by $n$ sensors:
\begin{align*}
x^{(i)}_{k+1} = A_ix^{(i)}_{k} + w^{(i)}_k,\\
y^{(i)}_k = C_ix^{(i)}_{k} + v^{(i)}_k,
\end{align*}
where $i \in N \triangleq \{1,\ldots,n\}$, $x_k^{(i)}\in\mathbb{R}^{n_i}$ is the state of the $i$-th process at time $k$ and $y_k^{(i)}\in\mathbb{R}^{m_i}$ is the measurement at time $k$. The state disturbance noise $w_k^{(i)}$'s, the measurement noise $v_k^{(i)}$'s and the initial state $x_0^{(i)}$'s of each process are mutually independent Gaussian random variables with distribution $\mathcal{N}(0,Q_i)$, $\mathcal{N}(0, R_i)$ and $\mathcal{N}(0, \Pi_i)$, respectively. We assume that $Q_i$'s and $\Pi_i$'s are positive semidefinite, and $R_i$'s are positive definite. In this work, we consider the case when all the processes are unstable \footnote{The structural results for stable processes can also be obtained in the MDP framework, but the proofs are more lengthy. We consider the unstable processes for brevity.}. We further assume that $(A_i,\sqrt{Q_i})$ is stabilizable and $(A_i, C_i)$ is detectable for each process.

The sensors are assumed to have sufficient computation capacity. After obtaining the measurement $y_k^{(i)}$, the sensors calculate the local minimum mean square error (MMSE) estimate of $x_k^{(i)}$ using a Kalman filter:
\begin{align*}
\hat{x}_{local,k}^{{(i)}-} &= A_i\hat{x}_{local,k-1}^{(i)},\\
P_{local,k}^{{(i)}-} &= A_iP_{local,k-1}^{(i)}A_i^T + Q_i,\\
K_{local,k}^{(i)} &= P_{local,k}^{{(i)}-}C_i^T(C_iP_{local,k}^{{(i)}-}C_i^T+R_i)^{-1},\\
\hat{x}_{local,k}^{(i)} &= \hat{x}_{local,k}^{{(i)}-} + K_{local,k}^{(i)}(y_k^{(i)}-C_i\hat{x}_{local,k}^{{(i)}-}),\\
P_{local,k}^{(i)} &= (I_{n_i}-K_{local,k}^{(i)}C_i)P_{local,k}^{{(i)}-},
\end{align*}
where $P_{local,k}^{{(i)}-}$ stands for the \emph{a prior} estimation error covariance, $P_{local,k}^{(i)}$ stands for the \emph{a posterior} estimation error covariance, and $K_{local,k}^{(i)}$ is the optimal gain. The iteration starts with $\hat{x}_{local,0}^{{(i)}-}=0$ and $P_{local,0}^{{(i)}-}=\Pi_i$. Since we assume that the initial error covariance matrix $\Pi_i$'s are positive semidefinite, $(A_i,\sqrt{Q_i})$ is stabilizable and $(A_i, C_i)$ is detectable for each process, the above iteration of the \emph{a posterior} estimation covariance $P_{local,k}^{(i)}$ converges exponentially fast to a steady state $\overline{P}^{(i)}$ \citep{anderson1979optimal}. Without loss of generality, the local estimation covariances are assumed to enter the steady state in our problem.

Because the communication bandwidth is limited, at each time step, only $m$ out of the $n$ sensors are allowed to send their local estimates, $\hat{x}_{local,k}^{(i)}$, to the remote estimator. Let $\gamma_k^{(i)} \in \{0, 1\}$ denote whether the $i$-th sensor transmits its estimate at time $k$, i.e., if $\hat{x}_{local,k}^{(i)}$ is sent to the remote estimator, $\gamma_k^{(i)}=1$; otherwise, $\gamma_k^{(i)}=0$. Denote $\theta=\{\gamma_k^{(i)}\}:i=1,2,\ldots,n;k=0,1,2,\ldots$ as the scheduling policy which specifies the value of $\gamma_k^{(i)}$ for each sensor at each time step.

%As the communication channel may not be reliable, we further assume that the packet arrivals of each packet are independent identically distributed random variables. Let $\alpha_k$ denote whether the transmission at time $k$ is successful or not.

Furthermore, as the sizes, $n_i$, of each dynamic process are different, the packet lengths of local estimate to be transmitted are also different. In some applications, e.g., underwater vehicles, the data rates are relatively low \citep{cui2006challenges}, and the local estimate is thus required to be split into more than one packet. The remote estimator needs all the relevant packets to decode the transmitted local estimate, $\hat{x}_{local,k}^{(i)}$. The local estimate of a higher dimension state may be split into more pieces than a lower dimension process. As a result, it may takes more time steps to transmit the whole packet for the higher dimensional process. Let $d_i$ denote the total time steps for the $i$-th process to transmit its local estimate. For example, suppose there are two processes. One of them is a scalar process whose dimension is $1$, while the dimension of the other is $2$. Suppose $d_1=1$ and $d_2=2$, then it takes $1$ time step for the first process to transmit its local estimate, and $2$ time steps for the second one.

For sensor $i$ at time $k$, denote $\eta_{k,\ell}^{(i)}=1$ for the arrival of $\hat{x}_{local,\ell}^{(i)}$ and $\eta_{k,\ell}^{(i)}=0$ otherwise. Define the time duration of missing local estimate at the remote estimator for the $i$-th sensor at time $k$ as
\begin{align}\label{eq:definition of tau}
\tau_k^{(i)} = k-\max_{k'}\{\ell : \eta_{k',\ell}^{(i)}=1\}.
\end{align}

Based on above settings, the remote estimator updates its remote estimation as follows:
\begin{align*}
\hat{x}_k^{(i)}=
    \begin{cases}
    A_i^{\tau_k^{(i)}} \hat{x}_{local,k-\tau_k^{(i)}}^{(i)}, &\text{if~} \eta_{k,\ell}^{(i)}=1,\\
    A_i \hat{x}_{k-1}^{(i)}, &\text{if~} \eta_{k,\ell}^{(i)}=0.
    \end{cases}
\end{align*}
As the local error covariance is assumed to have entered the steady state, the estimation covariance matrices at the remote estimator are as follows:
\begin{align*}
P_k^{(i)}=
    \begin{cases}
    h_i^{\tau_k^{(i)}} ( \overline{P}^{(i)} ), &\text{if~} \gamma_k^{(i)}=1,\\
    h_i(P^{(i)}_{k-1}), &\text{if~} \gamma_k^{(i)}=0,
    \end{cases}
\end{align*}
where the affine mapping of symmetric matrices $h_i^{\ell}(\cdot)$ and $h_i(\cdot)$ are defined as
\begin{align*}
&h_i^0(X) \triangleq X, \\
&h_i^{\ell}(X) \triangleq \underbrace{h_i \circ h_i \circ \cdots \circ h_i}_{\ell} (X), \\
&h_i(X) \triangleq A_i X A_i^T + Q_i,
\end{align*}
where $\circ$ denotes function composition. The following properties of $h_i(\cdot)$ will be useful in later sections.

\begin{lemma} (Lemma 3.1 in \cite{shi2012scheduling}) \label{lem:preliminary}
The Lyapunov-like operator $h_i^{\ell}(X)$ is monotonic with respect to $\ell$, i.e., $\forall i \in N$, if $\ell_1\leq\ell_2$ for $\ell_1,~\ell_2 \in \mathbb{Z}_+$, $h_i^{\ell_1}(\overline{P}^{(i)})\leq h_i^{\ell_2}(\overline{P}^{(i)})$. Consequently, $\forall \ell \in \mathbb{Z}_+$, $\Tr(\overline{P}^{(i)})<\Tr(h(\overline{P}^{(i)}))<\cdots<\Tr(h^{\ell}(\overline{P}^{(i)}))$.
\end{lemma}

The average per-stage cost of a scheduling policy is given as
\begin{align*}
J(\tau^{(1)}_0,\dots,\tau^{(n)}_0,\theta) \triangleq \limsup_{T\rightarrow \infty} \frac{1}{T} \mathbb{E} \Bigg[ \sum_{k=0}^{T-1} \sum_{i=1}^{n} \Tr(P_k^{(i)}) \Bigg].
\end{align*}

Note that the expectation is introduced because the scheduling policy might be random although the packet arrival is deterministic. With the above definition, the optimal scheduling policy is a feasible policy minimizing the total cost $J(\theta)$:
\begin{problem}\label{prb:problem1}
\begin{align*}
&\min_\theta J(\tau^{(1)}_0,\dots,\tau^{(n)}_0,\theta) \nonumber\\
s.t. ~&\sum_{i=1}^n \gamma_k^{(i)} = m, ~\forall k\geq0.
\end{align*}
\end{problem}

%\begin{rem}
%In this work, we do not consider the lossy network. This is because of the uncertainty of transmission decision in face of packet loss. Suppose it takes $\texttt{s}$ packets to transmit the local estimate, and when transmission of the $\texttt{r}$-th fails, it is not clear whether it should start transmitting the most updated estimate by abandoning the current estimate or not. This is left as future work.
%\end{rem}

\section{Main Results}
In this section, we solve Problem \ref{prb:problem1} by formulating it as an infinite time horizon Markov decision process with average cost criterion. First, we find that, without loss of any performance, the optimal scheduling policy can be restricted to be deterministic and stationary (independent of time index $k$). Furthermore, we show that the optimal policy is consistent and there is a threshold structure in the optimal policy.

%%%%%%%%%%%%%%%%%%%%%%%%%%%%%%%%%%%%%%%%%%%%%%%%%%%%%%%%%%%%%%%%%%%%%%%%%%%%%%%%%%%
\subsection{MDP Formulation}
For brevity, we assume $n=2$ and $m=1$ in the following discussion. We remark in Remark \ref{rem:general_case} that the structural results can be extended to general $n$ and $m$.

We formulate the Problem \ref{prb:problem1} as a discrete time MDP by a quadruplet $(\mathbb{S},\mathbb{A},P(\cdot|\cdot,\cdot),c(\cdot,\cdot))$. Each item above is explained as follows.

1) The state space $\mathbb{S}$ consists possible states, which are defined as $s \triangleq (\tau_1,\tau_2,\nu_1,\nu_2) \in \mathbb{N}^4$. The estimate holding time $\tau_1$ and $\tau_2$ are short notations for $\tau^{(i)}_k$ defined in \eqref{eq:definition of tau}, where the time index $k$ is omitted for brevity. The remaining packets counter $1\leq\nu_i\leq d_i$ indicates how many packets are left to be sent to complete the transmission of the estimate of the $i$-th sensor. For example, suppose the estimate of sensor $1$ consists of three packets and $\nu_1=2$ means that the first packet has been transmitted and there are still two packets to be transmitted.

2) The action $a$ is in the action space $\mathbb{A} \triangleq \{1, 2\}$, where $a=1$ or $2$ means that the first or the second sensor is scheduled, respectively. Note that the available action can be a random policy on $a=1$ and $a=2$ which also meets the constraint in Problem \ref{prb:problem1}.

3) The state transition, $P(s'|s,a)$, defines how the state transits from $s$ to $s'$ under action $a$. Because the channel is not lossy, the state transition under an action is deterministic. We state the state transition law as follows.

Given $s=(\tau_1,\tau_2,\nu_1,\nu_2)$ and $a=1$, the next state is
    \begin{align}
    &s'=\nonumber\\&
        \begin{cases}
        (1,\tau_2+1,1,1), &\text{if~} d_1=1 \\
        (\tau_1+1,\tau_2+1,\nu_1-1,d_2), &\text{if~} d_1 > 1 \text{~and~} \nu_1-1>0, \\
        (d_1,\tau_2+1,d_1,d_2), &\text{if~} d_1 > 1 \text{~and~} \nu_1-1=0.
        \end{cases}\label{eq:transition law 1}
    \end{align}
    The state transition from $s$ under $a=2$ is similar and omitted here.

4) The one-stage cost only depends on the current state and is defined as
\begin{align*}
c(s,a) \triangleq \Tr(h_1^{\tau_1}(\overline{P}^{(1)}))+\Tr(h_2^{\tau_2}(\overline{P}^{(2)})).
\end{align*}

%\begin{rem}
%Since $c(s,a)= \Tr(h_1^{\tau_1}(\overline{P}^{(1)}))+\Tr(h_2^{\tau_2}(\overline{P}^{(2)}))$, the cost only depends on the value of $\tau_1$ and $\tau_2$. Therefore, if there is no ambiguity, we would write $c(s,a)=c(\tau_1,\tau_2)$ for short.
%\end{rem}

Let $H_k=(s_0,a_0,\dots,s_{k-1},a_{k-1},s_k)$ be the history up to time $k$. A policy is a sequence $\{\pi_k\}_{k=0}^{\infty}$, where $\pi_k$'s are stochastic kernels from $H_k$ to $\mathbb{A}$. Denote $\Pi$ as the set of all such feasible policies. Define the average cost associated with an initial state $s_0$ and a policy $a_k=\pi_k(H_k)$ by
\begin{align*}
J(s_0,\{\pi_k\}_{k=0}^{\infty})=\limsup_{T\to\infty} \frac{1}{T}\mathbb{E}_{s_0}^{\{\pi_k\}_{k=0}^{\infty}}\Bigg[\sum_{k=0}^{T-1} c(s_k,\pi_k(s_k))\Bigg].
\end{align*}
One may see that the Problem \ref{prb:problem1} is equivalent to the following problem.
\begin{problem}
Find the optimal policy $\{\pi^*_k\}_{k=0}^{\infty}\in\Pi$ such that
\begin{align*}
J(s_0,\{\pi^*_k\}_{k=0}^{\infty})=\inf_{\{\pi_k\}_{k=0}^{\infty}\in\Pi}J(s_0,\{\pi_k\}_{k=1}^{\infty}).
\end{align*}
\end{problem}

%%%%%%%%%%%%%%%%%%%%%%%%%%%%%%%%%%%%%%%%%%%%%%%%%%%%%%%%%%%%%%%%%%%%%%%%%%%%%%%%%%%
\subsection{Structural Policy}
We first show that the optimal policy for the above MDP is deterministic and stationary, and satisfies the average optimality equality (AOE). A policy is deterministic and stationary if the policy depends only on the current state and there is no randomness in the actions. We write $\pi$ to denote the deterministic and stationary mapping from $\mathbb{S}$ to $\mathbb{A}$. Let $\Pi^{DS}$ be the set of all deterministic and stationary policies. The proof can be done by showing that the state space is finite if the optimal policy is adopted. Hence, we first prove the following lemma as a preliminary.

\begin{lemma}
The state space of the defined MDP problem is finite.
\end{lemma}
\textbf{Proof:}
We prove this result by constructing a Markov chain. Without loss of generality, assume $\d_1=\d_2=1$ as it shall be clear that the same construction can be done in other cases. Therefore, in this proof, we omit the last two components of the states and write $(\tau_1,\tau_2)$ to represent the state

%\textcolor{blue}{Note that when there is a stable process, the sensor measuring the stable process could never be scheduled. This is because the trace of the error covariance of the stable system is bounded. If such bound is less than the trace of the steady state error covariance of an unstable system, the unstable system should be scheduled in every time step. Therefore, there is no need to design a scheduling policy.}

\begin{figure}[t]
\centering
\begin{tikzpicture}[->, >=stealth', auto, semithick, node distance=1.3cm]
\node[state]                               (1) {(0,1)};
\node[state,right=of 1]                    (2) {(2,0)};
\node[state,right=of 2]                    (3) {(3,0)};
\node[draw=none,right=of 3]                (4) {$\ldots$};
\node[state,below=of 1]                    (5) {(1,0)};
\node[state,right=of 5]                    (6) {(0,2)};
\node[state,right=of 6]                    (7) {(0,3)};
\node[draw=none,right=of 7]                (8) {$\ldots$};

\draw[
    >=latex,
%   every node/.style={above,midway},% either
    auto=right,                      % or
    loop above/.style={out=75,in=105,loop},
    every loop,
    ]
     (1)   edge[bend right]         node {$2$} (5)
           edge[bend left]         node {$1$} (6)
     (2)   edge[bend right]        node {$1$} (1)
           edge                     node {$2$} (3)
     (3)   edge[bend right]        node {$1$} (1)
           edge                     node {$2$} (4)
     (5)   edge[bend right]        node {$1$} (1)
           edge[bend right]         node {$2$} (2)
     (6)   edge[bend left]        node {$2$} (5)
           edge                     node {$1$} (7)
     (7)   edge[bend left]        node {$2$} (5)
           edge                     node {$1$} (8);

\end{tikzpicture}

\caption{State transition diagram for the two sensors case.}\label{fig:markov chain}
\end{figure}
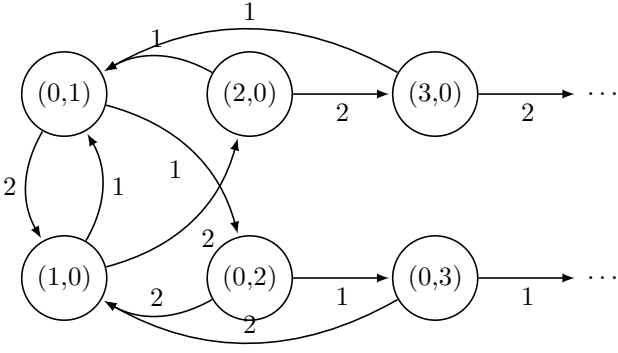

Observe that, when the time goes to infinity, the possible states can only be as follows
\begin{align*}
(0,1),~(1,0),~(0,2),~(2,0),~(0,3),~(3,0),\dots,
\end{align*}
i.e., either $\tau_1=0$ or $\tau_2=0$. This is because any state not in the above set will enter the possible set in one step. According to the transition law, the state transition diagram can then be constructed as in the Fig. \ref{fig:markov chain}.

There exists at least one policy that can enforce the number of the states in this Markov chain to be finite. For any infinite chain, we are always able to construct a sub-chain with finite states so that the average cost for the finite chain is smaller than the infinite chain.

When the state in the Markov chain is at infinity, the corresponding cost is infinity, i.e., $\displaystyle \lim_{\tau_1\to\infty} c(\tau_1,0)\geq \text{c}\rho^{2\tau_1}(A_i)>\infty$ for some constant $\text{c}>0$. Then a policy yielding the total number of nodes in the chain finite will render a finite average cost. Therefore, any policy yielding infinite chain is not optimal.

%Suppose $A_i$ is stable. Denote the number of nodes in the upper half of the chain in \ref{fig:markov chain} as $N_1$ and the lower half $N_2$, respectively. Then the average cost in this chain can be represented by
%\begin{align*}
%J(s,\pi)=\frac{1}{N_1}\sum_{k=2}^{N_1}\Tr[h_1^k(P^1)]+\frac{1}{N_2}\sum_{k=2}^{N_2}\Tr[h_1^k(P^2)].
%\end{align*}
%If the chain is infinite, which means that either $N_1$ or $N_2$ goes to infinity or both of them go to infinity, the corresponding cost is bounded by any pair of finite $N_1$ and finite $N_2$. The results follows from Lemma \ref{lem:preliminary}, which implies that the operator $h_i^{\ell}(X)$ is monotonic nondecreasing with respect to $\ell$.

Therefore, the optimal policy, if exists, should yields a finite Markov chain for the problem.
$\hfill \blacksquare$

The above lemma shows that we can restrict the idle durations of all sensors to be finite without loss of any performance. As both the state space and the action space are finite, we are able to verify that the MDP yields a deterministic and stationary optimal policy. We formally state the result in the following theorem.

\begin{theorem}
There exists a constant $\rho^*$, a continuous function $V(\cdot)$ on $\mathbb{S}$ and a deterministic and stationary policy $\pi^*\in\Pi^{DS}$ that satisfies the following AOE
\begin{align}\label{eq:ACOE}
\rho^*+V(s)=\min_{\pi\in\Pi^{DS}}\Big[ c(s,\pi(s))+\mathbb{E}^{\pi}_s[V] \Big],
\end{align}
where $\mathbb{E}^{\pi}_s[V]$ is the conditional expectation of the value of the next state under policy $\pi$, i.e., $\mathbb{E}^{\pi}_s[V]=\sum_{s'\in\mathbb{S}}V(s')P(s'|s,\pi)$, which in our problem reduces to $V(s')$.
\end{theorem}
\textbf{Proof:}
The results follows from Theorem 8.4.3 in \cite{puterman1994markov} as the state space and the action space is finite, the model is unichain and the cost is bounded.
$\hfill \blacksquare$

\begin{theorem}\label{thm:consistency}
Consistency. If $a^*_k=\pi^*(s)=i$ and $\nu_i\neq1$, $i=1,2$ , then $a^*_{k:k+\nu_i-1}=i$, where $a^*_{k:k+\nu_i-1}=i$ denotes $a^*_{k}=a^*_{k+1}=\cdots=a^*_{k+\nu_i-1}=i$.
\end{theorem}
\textbf{Proof:}
The proof can be found in the Appendix.
$\hfill \blacksquare$

\begin{remark}
Theorem \ref{thm:consistency} agrees with intuition. Theorem \ref{thm:consistency} states that once one chooses to schedule one sensor, he should not abort the scheduling halfway, i.e., he should wait for the completion of transmission of the whole estimate before choosing another sensor to schedule again. In other words, when the last transmission is done, the scheduler chooses a sensor to schedule. The action should be consecutively executed until the packet has been fully delivered.
\end{remark}

Now we present the threshold type policy of the MDP. The structural result will simplify off-line computation and facilitate online application of the scheduling policy. Note that we only consider the case when $\nu_1=d_1$ and $\nu_2=d_2$. As it is proved that the scheduling policy should be consistent, the decision of which sensor to schedule is only made when the packet(s) of one local estimate has(have) been fully transmitted.

\begin{theorem}\label{thm:threshold}
Threshold type policy.
\begin{enumerate}
\item If $\pi^*((\tau_1,\tau_2,d_1,d_2))=1$, then $\pi^*((\tau_1+z,\tau_2,d_1,d_2))=1$, where $z$ is any positive integer;
\item If $\pi^*((\tau_1,\tau_2,d_1,d_2))=2$, then $\pi^*((\tau_1,\tau_2+z,d_1,d_2))=2$, where $z$ is any positive integer.
\end{enumerate}
\end{theorem}
\textbf{Proof:}
The proof can be found in the Appendix.
$\hfill \blacksquare$

\begin{remark}\label{rem:general_case}
The threshold structure can be extended to general $m$ and $n$. Let $\nu_i=1, \forall i \in [1,n]$. For $1\leq i \leq n$, define $\tau_i^-=(\tau_1,\ldots,\tau_{i-1},\tau_{i+1},\ldots,\tau_{n})$ as the state of the system except for the $i$-th process. There exists measurable functions $\phi_i:\mathbb{N}^{n-1}\mapsto\mathbb{N}$ and the optimal policy is as follows:
\begin{enumerate}
\item Choose the $i$-th sensor if $\phi_i(\tau_i^-)\leq\tau_i$;
\item Don't choose the $i$-th sensor if $\phi_i(\tau_i^-)>\tau_i$.
\end{enumerate}
\end{remark}

The threshold policy has valuable properties because only the state on the boundary should be stored for implementation. When the scheduler needs to schedule a sensor, only the comparison between the current state and the boundary is needed. This reduces the space required in online implementation. Furthermore, if one knows the threshold structure, he can develop special algorithms for policy iteration to reduce the spatial complexity of the MDP. Further discussion can be found in section $4.7$ of \cite{puterman1994markov}.

%%%%%%%%%%%%%%%%%%%%%%%%%%%%%%%%%%%%%%%%%%%%%%%%%%%%%%%%%%%%%%%%%%%%%%%%%%%%%%%%%%%
\section{NUMERICAL EXAMPLES}
In this section, we present a numerical example to illustrate our theoretical results, namely the consistency (Theorem \ref{thm:consistency}) and the threshold policy (Theorem \ref{thm:threshold}).

Let $n=2$ and $m=1$ and the parameters of the two processes be as follows
\begin{gather*}
A_1=1.4, ~C_1=1, ~Q_1=1, ~R_1=1;\\
A_2=\begin{bmatrix}1.2&1\\0&1\end{bmatrix}, ~C_2=[1~0],
Q_2=\begin{bmatrix}1&0\\0&1\end{bmatrix}, ~R_2=1.
\end{gather*}
Moreover, the packet length of the first process is $3$, while the second $4$. It can be seen that the spectral radius are $\rho(A_1)=1.4$ and $\rho(A_2)=1.2$, respectively. By solving the corresponding Riccati equations, the steady state error covariance matrices at the sensor side are
\begin{align*}
\overline{P}^{(1)}=0.70, \overline{P}^{(2)}=\begin{bmatrix}0.84&0.40\\0.40&2.00\end{bmatrix}.
\end{align*}
We use the MDP toolbox \citep{chades2009markov} to calculate the optimal policy through the value iteration algorithm.

\begin{figure}
    \includegraphics[width=0.4\textwidth]{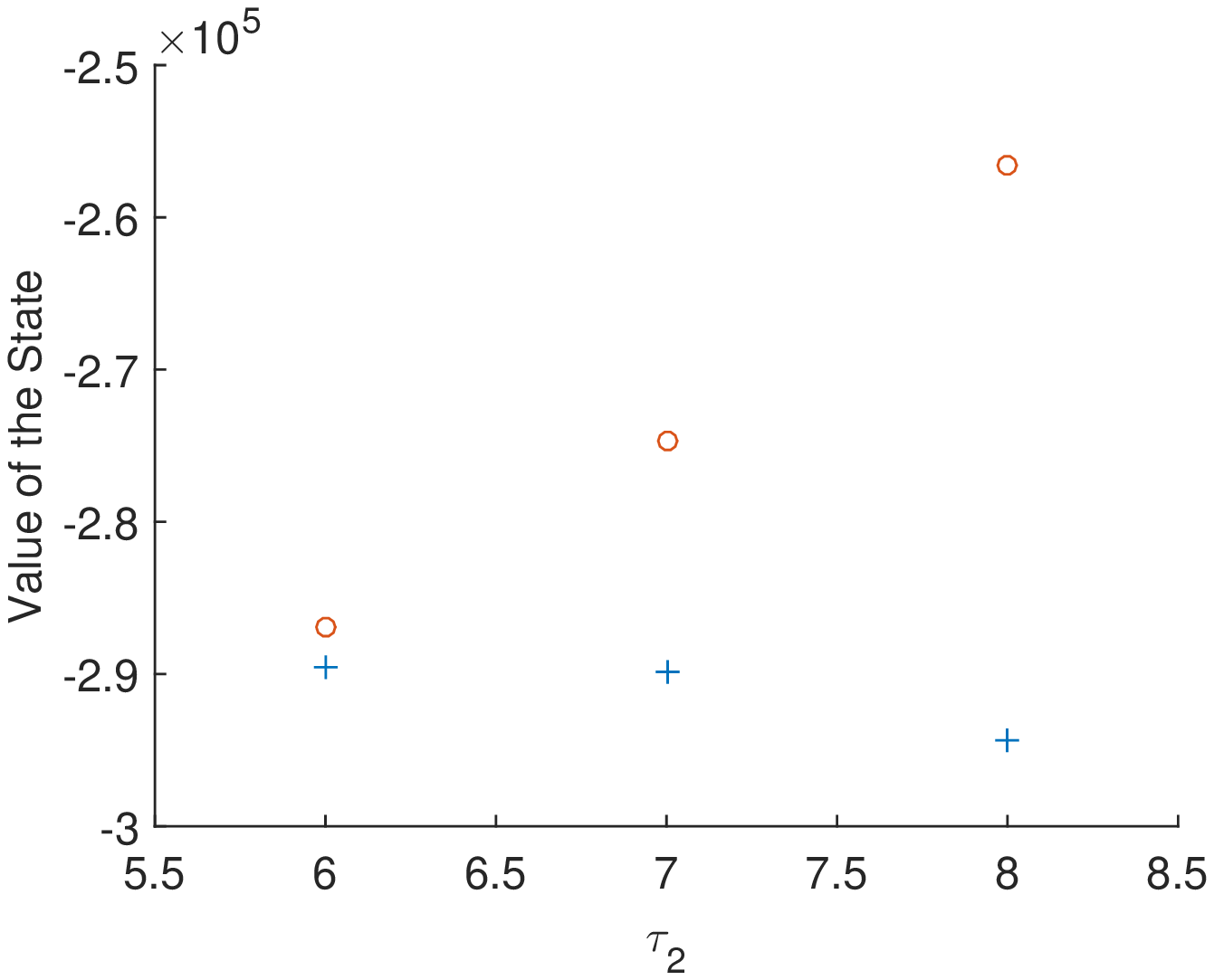}
    \caption{Visualization of consistency in Theorem \ref{thm:consistency}. The meaning of adopted mark is the same as Fig. \ref{fig:verify_threshold}.}
    \label{fig:consistency}

    \includegraphics[width=0.4\textwidth]{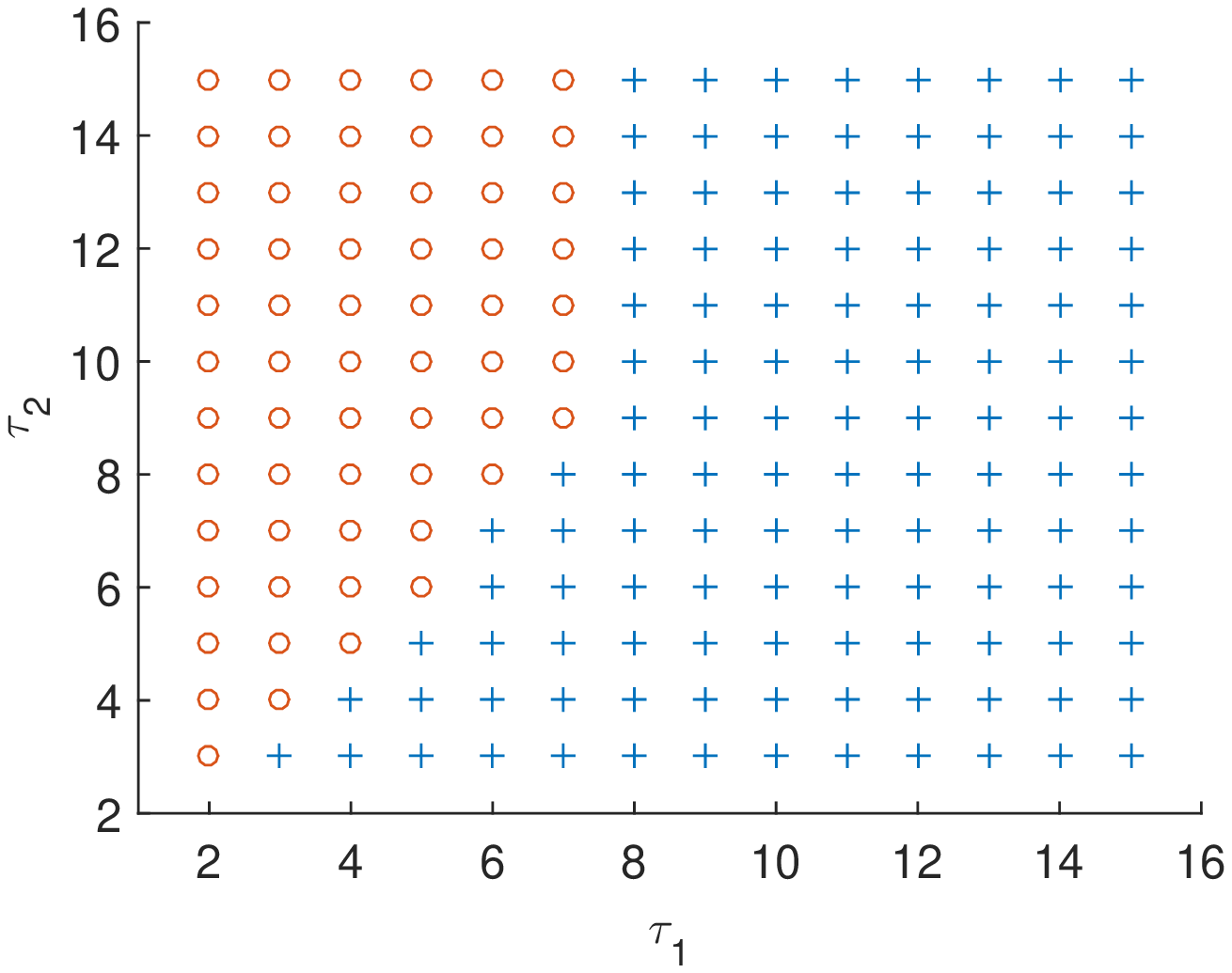}
    \caption{Visualization of threshold policy. The blue $'+'$ sign stands for scheduling sensor 1, and red circle sign for sensor 2.}
    \label{fig:threshold}

    \includegraphics[width=0.4\textwidth]{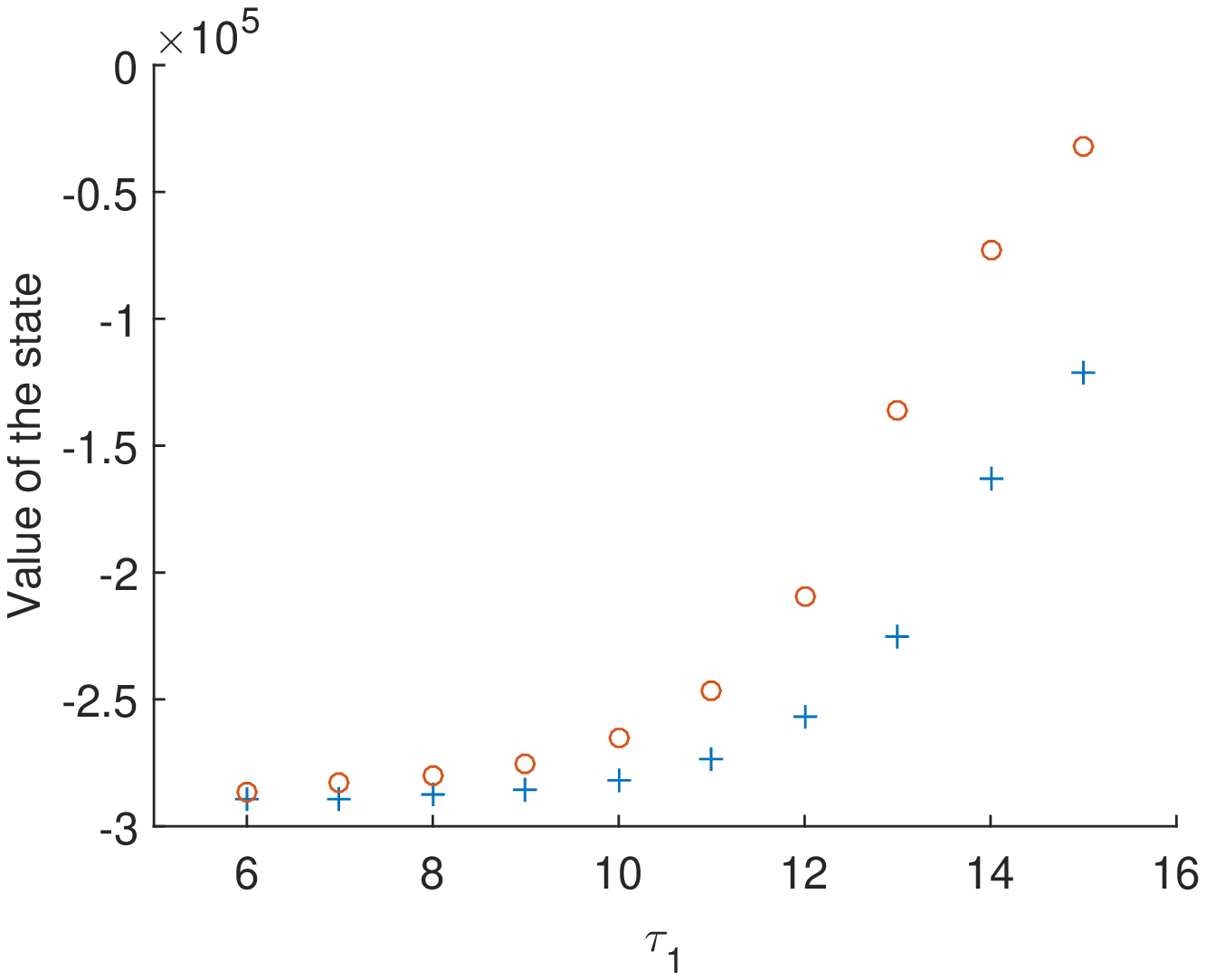}
    \caption{Illustration of threshold policy. The blue '+' sign stands for the value of the future state if sensor 1 is chosen, and red circle sign for sensor 2.}
    \label{fig:verify_threshold}
\end{figure}

Fig.\ref{fig:consistency} illustrates the consistency of optimal policy in Theorem \ref{thm:consistency}. According to the result of value iteration, sensor 1 should be scheduled at $(6,6,3,4)$. Because of consistency of the optimal policy, the transmission of the local estimate should not be interrupted. To complete the transmission, sensor $1$ should continue its transmission at $(7,7,2,4)$ and $(8,8,1,4)$. Accordingly, the next states are $(7,7,2,4)$, $(8,8,1,4)$ and $(2,9,3,4)$ in order. Fig. \ref{fig:consistency} presents the value of the possible `next' state if either sensor $1$ or $2$ is chosen to schedule. The figure shows that scheduling sensor $1$ always cause the value of next state to be less than scheduling sensor $2$.

The threshold policy is presented in Fig. \ref{fig:threshold}. It is clear that there is a decision boundary on the plane. Note that at state $(6,6,3,4)$, it is optimal to schedule sensor $1$. Following from the threshold policy, it is also optimal to schedule sensor 1 for $(x,6,3,4), x\geq6$. By fixing $\tau_2=6,~\nu_1=\nu_2=1$, Fig. \ref{fig:verify_threshold} shows the value of the next state of the system if either sensor $1$ or $2$ is scheduled. We can observe that scheduling sensor $1$ always yields the next state less valuable.

%%%%%%%%%%%%%%%%%%%%%%%%%%%%%%%%%%%%%%%%%%%%%%%%%%%%%%%%%%%%%%%%%%%%%%%%%%%%%%%%%%%
\section{CONCLUSION}

This paper has considered sensor scheduling problem in a time-slotted setting. Depending on the size of the packet to be transmitted, different time slots are allocated to different sensors. We have studied the problem of minimizing the average estimation covariance under the constraints on packet length and bandwidth, and have formulated it as an infinite time horizon Markov decision process with average cost criterion. We have found that there is a deterministic and stationary policy for the problem. Furthermore, we have shown that the optimal policy has consistency and threshold structures. The consistent behavior means that once a sensor is chosen to schedule, the transmission of current estimate should not be interrupted. The threshold type structure has reduced searching space and has facilitated online implementation. A numerical example has been provided to illustrate our results.

%%%%%%%%%%%%%%%%%%%%%%%%%%%%%%%%%%%%%%%%%%%%%%%%%%%%%%%%%%%%%%%%%%%%%%%%%%%%%%%%%%%
%\bibliographystyle{ifacconf}
%\bibliographystyle{elsarticle-harv}
%\bibliographystyle{ifacconf-harvard}
\bibliography{plreference}

\begin{thebibliography}{18}
\providecommand{\natexlab}[1]{#1}
\providecommand{\url}[1]{\texttt{#1}}
\providecommand{\urlprefix}{URL }
\expandafter\ifx\csname urlstyle\endcsname\relax
  \providecommand{\doi}[1]{doi:\discretionary{}{}{}#1}\else
  \providecommand{\doi}{doi:\discretionary{}{}{}\begingroup
  \urlstyle{rm}\Url}\fi

\bibitem[{Akyildiz et~al.(2002)Akyildiz, Su, Sankarasubramaniam, and
  Cayirci}]{akyildiz2002wireless}
Akyildiz, I.F., Su, W., Sankarasubramaniam, Y., and Cayirci, E. (2002).
\newblock Wireless sensor networks: a survey.
\newblock \emph{Computer networks}, 38(4), 393--422.

\bibitem[{Anderson and Moore(1979)}]{anderson1979optimal}
Anderson, B. and Moore, J.B. (1979).
\newblock Optimal filtering.
\newblock \emph{Englewood Cliffs: Prentice-Hall}.

\bibitem[{Chad{\`e}s et~al.(2009)Chad{\`e}s, Cros, Garcia, and
  Sabbadin}]{chades2009markov}
Chad{\`e}s, I., Cros, M., Garcia, F., and Sabbadin, R. (2009).
\newblock Markov decision processes ({MDP}) toolbox.
\newblock \emph{URL http://www.inra.fr/mia/T/MDPtoolbox/}.

\bibitem[{Chamam and Pierre(2007)}]{chamam2007optimal}
Chamam, A. and Pierre, S. (2007).
\newblock Optimal scheduling of sensors' states to maximize network lifetime in
  wireless sensor networks.
\newblock In \emph{2007 IEEE International Conference on Mobile Adhoc and
  Sensor Systems}, 1--6. IEEE.

\bibitem[{Cui et~al.(2006)Cui, Kong, Gerla, and Zhou}]{cui2006challenges}
Cui, J.H., Kong, J., Gerla, M., and Zhou, S. (2006).
\newblock The challenges of building mobile underwater wireless networks for
  aquatic applications.
\newblock \emph{Ieee Network}, 20(3), 12--18.

\bibitem[{Gupta et~al.(2006)Gupta, Chung, Hassibi, and
  Murray}]{gupta2006stochastic}
Gupta, V., Chung, T.H., Hassibi, B., and Murray, R.M. (2006).
\newblock On a stochastic sensor selection algorithm with applications in
  sensor scheduling and sensor coverage.
\newblock \emph{Automatica}, 42(2), 251--260.

\bibitem[{Han et~al.(2017)Han, Wu, Zhang, and Shi}]{han2017optimal}
Han, D., Wu, J., Zhang, H., and Shi, L. (2017).
\newblock Optimal sensor scheduling for multiple linear dynamical systems.
\newblock \emph{Automatica}, 75, 260--270.

\bibitem[{Hern{\'a}ndez-Lerma and Lasserre(1999)}]{hernandez2012further}
Hern{\'a}ndez-Lerma, O. and Lasserre, J.B. (1999).
\newblock \emph{Further topics on discrete-time {Markov} control processes},
  volume~42.
\newblock Springer.

\bibitem[{Hovareshti et~al.(2007)Hovareshti, Gupta, and
  Baras}]{hovareshti2007sensor}
Hovareshti, P., Gupta, V., and Baras, J.S. (2007).
\newblock Sensor scheduling using smart sensors.
\newblock In \emph{Decision and Control, 2007 46th IEEE Conference on},
  494--499. IEEE.

\bibitem[{Mo et~al.(2014)Mo, Garone, and Sinopoli}]{mo2014infinite}
Mo, Y., Garone, E., and Sinopoli, B. (2014).
\newblock On infinite-horizon sensor scheduling.
\newblock \emph{Systems \& control letters}, 67, 65--70.

\bibitem[{Palattella et~al.(2013)Palattella, Accettura, Grieco, Boggia, Dohler,
  and Engel}]{palattella2013optimal}
Palattella, M.R., Accettura, N., Grieco, L.A., Boggia, G., Dohler, M., and
  Engel, T. (2013).
\newblock On optimal scheduling in duty-cycled industrial {IoT} applications
  using {IEEE}802.15.4e {TSCH}.
\newblock \emph{IEEE Sensors Journal}, 13(10), 3655--3666.

\bibitem[{Puterman(1994)}]{puterman1994markov}
Puterman, M.L. (1994).
\newblock \emph{Markov decision processes: discrete stochastic dynamic
  programming}.
\newblock John Wiley \& Sons, Inc.

\bibitem[{Ren et~al.(2016)Ren, Wu, Dey, and Shi}]{ren2016attack}
Ren, X., Wu, J., Dey, S., and Shi, L. (2016).
\newblock Attack allocation on remote state estimation in multi-systems:
  structural results and asymptotic solution.
\newblock \emph{arXiv preprint arXiv:1609.00887}.

\bibitem[{Shi et~al.(2011)Shi, Cheng, and Chen}]{shi2011sensor}
Shi, L., Cheng, P., and Chen, J. (2011).
\newblock Sensor data scheduling for optimal state estimation with
  communication energy constraint.
\newblock \emph{Automatica}, 47(8), 1693--1698.

\bibitem[{Shi and Zhang(2012)}]{shi2012scheduling}
Shi, L. and Zhang, H. (2012).
\newblock Scheduling two {Gauss-Markov} systems: an optimal solution for remote
  state estimation under bandwidth constraint.
\newblock \emph{IEEE Transactions on Signal Processing}, 60(4), 2038--2042.

\bibitem[{Watteyne et~al.(2015)Watteyne, Palattella, and
  Grieco}]{watteyne2015using}
Watteyne, T., Palattella, M., and Grieco, L. (2015).
\newblock Using {IEEE} 802.15.4e time-slotted channel hopping ({TSCH}) in the
  internet of things ({IoT}): Problem statement.
\newblock Technical report.

\bibitem[{Wu et~al.(2013)Wu, Jia, Johansson, and Shi}]{wu2013event}
Wu, J., Jia, Q.S., Johansson, K.H., and Shi, L. (2013).
\newblock Event-based sensor data scheduling: trade-off between communication
  rate and estimation quality.
\newblock \emph{IEEE Transactions on automatic control}, 58(4), 1041--1046.

\bibitem[{Zhao et~al.(2014)Zhao, Zhang, Hu, Abate, and
  Tomlin}]{zhao2014optimal}
Zhao, L., Zhang, W., Hu, J., Abate, A., and Tomlin, C.J. (2014).
\newblock On the optimal solutions of the infinite-horizon linear sensor
  scheduling problem.
\newblock \emph{IEEE Transactions on Automatic Control}, 59(10), 2825--2830.

\end{thebibliography}

%%%%%%%%%%%%%%%%%%%%%%%%%%%%%%%%%%%%%%%%%%%%%%%%%%%%%%%%%%%%%%%%%%%%%%%%%%%%%%%%%%%
\appendix

\section{Proofs}

This appendix serves to provide the proof of Theorem \ref{thm:consistency} and \ref{thm:threshold} by analyzing the value function of the states. Because the average cost MDP problem involves a relative value function, analyzing the structure of the value of the states is difficult. Thanks to the existence of a deterministic and stationary policy, we can analyze the structure of the optimal policy in the average cost problem by analyzing the discounted cost function. The idea is similar to the proof of the Theorem $2$ in \cite{ren2016attack}. For $0<\alpha<1$, define the discounted cost under a deterministic and stationary policy $\pi$ and an initial state $s_0$ by
\begin{align}\label{eq:discounted cost problem}
J_{\alpha}(s_0,\pi) = \limsup_{T\to\infty} \frac{1}{T}\mathbb{E}_s^{\pi}\Bigg[\sum_{k=0}^{T-1}\alpha^k c(s_k,a_k)\Bigg].
\end{align}
Define $J_{\alpha}^*(s_0)=\inf_{\pi\in\Pi}J_{\alpha}(s_0,\pi)$ and $u(s)=J_{\alpha}^*(s)-J_{\alpha}^*(0,0,d_1,d_2)$. Because there exists an optimal policy to the average cost counterpart of the cost function \eqref{eq:discounted cost problem}, the limit of $u(s)$ as $\alpha$ goes to $1$ exits and is the relative value function in \eqref{eq:ACOE}, i.e., 
\begin{align*}
V(s)=\lim_{\alpha\uparrow1}u(s).
\end{align*}
With this observation, we can analyze the properties of $V(s)$ by examining $u(s)$ by value iteration. Define the dynamic programming operator $T_{\alpha}u(\cdot)$ for a given measurable function $u:~\mathbb{S}\mapsto\mathbb{R}$ as
\begin{align}\label{eq:dynamic operator}
T_{\alpha}u(s)\triangleq\min_{a\in\Pi^{DS}}\Big[c(s,a)+\alpha \mathbb{E}^{\pi}_s[u] \Big], ~s\in\mathbb{S}.
\end{align}
It follows from \cite{hernandez2012further} that $T_{\alpha}u(s)$ is a contraction mapping. By Banach fixed-point theorem,
\begin{align*}
\lim_{n\to\infty}T_{\alpha}^n u(s)=J_{\alpha}^*(s),
\end{align*}
which guarantees that certain properties holds for $u(s)$ if the dynamic operator preserves such properties. Briefly speaking, thanks to the above observations, we can obtain the structure of $V(s)=\lim_{\alpha\uparrow1}u(s)$ by verifying that the dynamic operator \eqref{eq:dynamic operator} preserves the same structure.

Before giving the proofs of Theorem \ref{thm:consistency} and \ref{thm:threshold}, we need the following lemma to reveal the monotonicity structure of the value function $V(s)$ in the discounted cost MDP.

Because the cost $c(s,a)$ only depends on $\tau_1$ and $\tau_2$, and $V(s)$ and $u(s)$ depends of $\tau_1$, $\tau_2$, $\nu_1$ and $\nu_2$, we write $c(s)$ as $c(\tau_1,\tau_2)$, $u(s)$ as $u(\tau_1,\tau_2,\nu_1,\nu_2)$ and $V(s)$ as $V(\tau_1,\tau_2,\nu_1,\nu_2)$.

\begin{lem}\label{lem:monotone_value_fn}
Monotonicity.
The value function $V(s)$ of the states $s$ is monotonic with respect to the first two variables, respectively, i.e., $V(\tau_1,\cdot,\cdot,\cdot)\geq V(\tau'_1,\cdot,\cdot,\cdot)$ if $\tau_1\geq\tau'_1$ and $V(\cdot,\tau_2,\cdot,\cdot)\geq V(\cdot,\tau'_2,\cdot,\cdot)$ if $\tau_2\geq\tau'_2$.
\end{lem}

\textbf{Proof:}
Suppose $\tau_1\geq\tau_1'$, $\tau_2\geq\tau_2'$, $\nu_1=\nu_1'$, $\nu_2=\nu_2'$, and $u(s)\geq u(s')$. Since $c(s)\geq c(s'), ~\forall a$, we have
\begin{align*}
c(s)+\alpha\mathbb{E}^{\pi}_s[u] \geq c(s')+\alpha\mathbb{E}^{\pi}_{s'}[u],
\end{align*}
which suggests that $T_{\alpha}u(s)\geq T_{\alpha}u(s')$. By the previous discussion of the contraction property of $T_{\alpha}u(s)$, the monotonicity of $V(s)$ is proven.
$\hfill \blacksquare$

\subsection{Proof of Theorem~\ref{thm:consistency}}

\textbf{Proof:}
Consider the case when $d_1>2$ and $d_2>2$. Other cases can be easily extended. Without loss of generality, assume that $s=(\tau_1,\tau_2,d_1,d_2)$ and it is optimal to schedule sensor $1$, i.e., $a^*=\pi^*(\tau_1,\tau_2,d_1,d_2)=1$. Note that $\nu_i=d_i$ implies that one transmission of a local estimate have been complete at $k-1$. As a result, a novel transmission of the local estimate of sensor $1$ starts at time $k$. This indicates that
\begin{align*}
&c(\tau_1,\tau_2)+\alpha u(\tau_1+1,\tau_2+1,d_1-1,d_2) \\
 < &c(\tau_1,\tau_2)+\alpha u(\tau_1+1,\tau_2+1,d_1,d_2-1).
\end{align*}
If $a^*_{k+1}=1$, then we have
\begin{multline*}
c(\tau_1,\tau_2)+\alpha u(\tau_1+1,\tau_2+1,d_1-1,d_2)\\
=c(\tau_1,\tau_2)+\alpha c(\tau_1+1,\tau_2+1)+\alpha^2 u(\tau_1+2,\tau_2+2,d_1-2,d_2).
\end{multline*}
Since we always have $c(\tau_1+1,\tau_2+1)>0$,
\begin{align*}
u(\tau_1+1,\tau_2+1,d_1-1,d_2) > u(\tau_1+2,\tau_2+2,d_1-2,d_2).
\end{align*}
If $a^*_{k+1}=2$, then
\begin{multline*}
c(\tau_1+1,\tau_2+1) + \alpha u(\tau_1+2,\tau_2+2,d_1-2,d_2) \\
> c(\tau_1+1,\tau_2+1)+\alpha u(\tau_1+2,\tau_2+2,d_1,d_2-1),
\end{multline*}
which suggests
\begin{align*}
u(\tau_1+2,\tau_2+2,d_1-2,d_2) > u(\tau_1+2,\tau_2+2,d_1,d_2-1).
\end{align*}
On the other hand, due to monotonicity, we have
\begin{align*}
u(\tau_1+2,\tau_2+2,d_1,d_2-1) > u(\tau_1+1,\tau_2+1,d_1,d_2-1).
\end{align*}
Hence,
\begin{align*}
u(\tau_1+2,\tau_2+2,d_1-2,d_2) > u(\tau_1+1,\tau_2+1,d_1,d_2-1),
\end{align*}
which causes contradiction.

Therefore, we obtain $a^*_{k+1}=1$. By mathematical induction and the contraction mapping argument, the proof is complete.
$\hfill \blacksquare$

\subsection{Proof of Theorem \ref{thm:threshold}}
\textbf{Proof:}
The Theorem is equivalent to the following:
\begin{enumerate}
\item If $V(\tau_1,\tau_2,d_1-1,d_2)\leq V(\tau_1,\tau_2,d_1,d_2-1)$, then $V(\tau_1+z,\tau_2,d_1-1,d_2)\leq V(\tau_1+z,\tau_2,d_1,d_2-1)$, where $z$ is any positive integer;
\item If $V(\tau_1,\tau_2,d_1-1,d_2)\geq V(\tau_1,\tau_2,d_1,d_2-1)$, then $V(\tau_1,\tau_2+z,d_1-1,d_2)\geq V(\tau_1,\tau_2+z,d_1,d_2-1)$, where $z$ is any positive integer.
\end{enumerate}
We prove the first case and the second case can be proved similarly. As it is done in the proof of Theorem \ref{thm:consistency}, we prove the structure of $V(\cdot)$ by showing that $u(\cdot)$ has the same structure.

Depending on whether the packet length $d_i$ is greater than $1$ or not, the proof is divided into four cases.

1) Case $d_1=d_2=1$.\\ $u(1,\tau_2+1,d_1,d_2)\leq u(\tau_1+1,1,d_1,d_2)\leq u(\tau_1+z+1,1,d_1,d_2)$. The last inequality is due to the monotonicity from Lemma \ref{lem:monotone_value_fn}.

2) Case $d_1=1$ and $d_2\neq 1$.\\ Because of monotonicity, we similarly have $u(1,\tau_2+1,d_1,d_2)\leq u(\tau_1+1,\tau_2+1,d_1,d_2-1)\leq u(\tau_1+z+1,\tau_2+1,d_1,d_2-1)$.

3) Case $d_1\neq1$ and $d_2\neq1$.\\
    If $u(\tau_1,\tau_2,d_1-1,d_2)\leq u(\tau_1,\tau_2,d_1,d_2-1)$, from Theorem \ref{thm:consistency}, we obtain
    \begin{multline*}
    c(\tau_1,\tau_2)+\alpha u(\tau_1+1,\tau_2+1,d_1-1,d_2) \\
    \leq c(\tau_1,\tau_2)+\alpha u(\tau_1+1,\tau_2+1,d_1,d_2-1),
    \end{multline*}
    which implies that
    \begin{align}\label{eq:implication}
    u(\tau_1+1,\tau_2+1,d_1-1,d_2) \leq u(\tau_1+1,\tau_2+1,d_1,d_2-1).
    \end{align}

    Assume $u(\tau+z,\tau_2,d_1-1,d_2) > u(\tau_1+z,\tau_2,d_1,d_2-1)$, then we have
    \begin{multline*}
    u(\tau_1+z,\tau_2,d_1-1,d_2) \\
    > c(\tau_1+z,\tau_2)+\alpha u(\tau_1+z+1,\tau_2+1,d_1,d_2-1),
    \end{multline*}
    which implies that
    \[u(\tau_1+z,\tau_2,d_1-1,d_2) > u(\tau_1+z+1,\tau_2+1,d_1,d_2-1).\]
    Meanwhile, note that
    \[u(\tau_1+z+1,\tau_2+1,d_1,d_2-1) > u(\tau_1+1,\tau_2+1,d_1,d_2-1),\]
    hence
    \[u(\tau_1+z,\tau_2,d_1-1,d_2) > u(\tau_1+1,\tau_2+1,d_1,d_2-1). \]
    However, due to \eqref{eq:implication}
    \[ u(\tau_1+1,\tau_2+1,d_1,d_2-1) \geq  u(\tau_1+1,\tau_2+1,d_1-1,d_2),\]
    which causes
    \[u(\tau_1+z,\tau_2,d_1-1,d_2) > u(\tau_1+1,\tau_2+1,d_1-1,d_2). \]
    This violate the monotonicity and consistency of the value function and hence the assumption is incorrect. Therefore, $u(\tau+z,\tau_2,d_1-1,d_2) \leq u(\tau_1+z,\tau_2,d_1,d_2-1)$.

4) Case $d_1\neq1$ and $d_2=1$.\\This is similar to the case $d_1\neq1$ and $d_2\neq1$ and is omitted here.
$\hfill \blacksquare$

\end{document}